\documentclass[runningheads]{llncs}
\usepackage[T1]{fontenc}
\usepackage{subcaption}
\usepackage{orcidlink}
\usepackage{hyperref}

\newcommand\submittedtext{%
\footnotesize \textbf{Preprint}. This work has been accepted to the 23rd International Conference on Service-Oriented Computing (ICSOC 2025). © 2025 Springer. Personal use of this material is permitted. Permission from Springer must be obtained for all other uses, in any current or future media.}

\newcommand\submittednotice{%
\begin{tikzpicture}[remember picture,overlay]
\node[anchor=south,yshift=10pt] at (current page.south) {\fbox{\parbox{\dimexpr1\textwidth-\fboxsep-\fboxrule\relax}{\submittedtext}}};
\end{tikzpicture}%
}

\usepackage{graphicx}
\usepackage{multirow}
\usepackage{xspace}
\usepackage{float}
\usepackage{amsfonts}  %
\usepackage{pifont}    %

\usepackage{cite}
\usepackage{amsmath,amssymb,amsfonts}
\usepackage{algorithmic}
\usepackage{graphicx}
\usepackage{textcomp}
\usepackage{xcolor}
\usepackage{xcolor}
\usepackage{cleveref}
\usepackage{comment}
\usepackage{booktabs}
\usepackage{atbegshi}
\usepackage{comment}

\newcommand{\hide}[1]{} %

\def\BibTeX{{\rm B\kern-.05em{\sc i\kern-.025em b}\kern-.08em
    T\kern-.1667em\lower.7ex\hbox{E}\kern-.125emX}}

\usepackage{listings}
\usepackage{xcolor}

\definecolor{codegreen}{rgb}{0,0.6,0}
\definecolor{codegray}{rgb}{0.5,0.5,0.5}
\definecolor{codepurple}{rgb}{0.58,0,0.82}
\lstdefinestyle{psla}{
    basicstyle=\ttfamily\scriptsize,
    keywordstyle=\color{blue}\bfseries,
    commentstyle=\color{codegreen},
    stringstyle=\color{codepurple},
    showstringspaces=false,
    breaklines=true,
    frame=single,
    captionpos=b,
    tabsize=2
}

\begin{document}

\title{Towards Trusted Service Monitoring: \\Verifiable Service Level Agreements}
\titlerunning{Towards Trusted Service Monitoring}
\author{Fernando Castillo\inst{1}\orcidlink{0009-0003-6835-8711} \and
Eduardo Brito\inst{2,3}\orcidlink{0009-0002-9996-6333} \and
Sebastian Werner\inst{1}\orcidlink{0000-0001-8051-7226} \and
Pille Pullonen-Raudvere\inst{2}\orcidlink{0000-0002-3255-7001} \and
Jonathan Heiss\inst{1}\orcidlink{0000-0002-4239-8534}}

\institute{
  Information Systems Engineering, Technische Universität Berlin, Berlin, Germany\\
  \email{\{fc,sw,jh\}@ise.tu-berlin.de}
  \and
   Cybernetica AS, Tallin, Estonia\\
  \email{\{eduardo.brito, pille.pullonen-raudvere\}@cyber.ee}
   \and
   University of Tartu, Tartu, Estonia
}

\authorrunning{F. Castillo et al.}

\maketitle
\submittednotice

\begin{abstract}

Service Level Agreement (SLA) monitoring in service-oriented environments suffers from inherent trust conflicts when providers self-report metrics, creating incentives to underreport violations. We introduce a framework for generating verifiable SLA violation claims through trusted hardware monitors and zero-knowledge proofs, establishing cryptographic foundations for genuine trustworthiness in service ecosystems. Our approach starts with machine-readable SLA clauses converted into verifiable predicates and monitored within Trusted Execution Environments. These monitors collect timestamped telemetry, organize measurements into Merkle trees, and produce signed attestations. Zero-knowledge proofs aggregate Service-Level Indicators to evaluate compliance, generating cryptographic proofs verifiable by stakeholders, arbitrators, or insurers in disputes, without accessing underlying data. This ensures three security properties: integrity, authenticity, and validity. Our prototype demonstrates linear scaling up to over 1 million events per hour for measurements with near constant-time proof generation and verification for single violation claims, enabling trustless SLA enforcement through cryptographic guarantees for automated compliance verification in service monitoring.

\end{abstract}

\section{Introduction}
\label{sec:introduction}

Service Level Agreements (SLAs) are widely used to codify non-functional guarantees in cloud and edge systems. Yet, verifying whether these guarantees are upheld remains problematic. SLA monitoring is often under the control of the provider or a third party~\cite{hussain2014maintaining}, creating trust asymmetries and limiting transparency~\cite{freitas2011cost}, especially in adversarial settings where providers may benefit from under-reporting, consumers may seek unjustified compensation, or competitors may launch reputation attacks~\cite{macias2016analysis}. %

This lack of verifiability also undermines broader accountability infrastructures in service ecosystems. In particular, cyber insurance, now a key instrument for managing cyber risk~\cite{novo2025applications}, faces significant challenges due to unverifiable claims of compliance, e.g., malfunctioning of a service or non-compliance of an SLA caused by a data breach. Studies report recurring disputes, delayed reimbursements, and difficulties in underwriting premiums, often due to self-reported or incomplete SLA evidence~\cite{aziz2020systematic,tsohou2023cyber}. Despite forecasts estimating the global cyber insurance market to exceed \$28 billion by 2026~\cite{tsohou2023cyber}, insurers lack cryptographic assurance over SLA fulfillment, complicating risk pooling and claims processing by needing to spend resources on verifying the authenticity, integrity and validity of the evidence used to generate claims, e.g, unrealized gains due to service unavailability on high spike selling days. %

To address this gap, we propose a framework for verifiable SLA compliance that leverages trusted hardware monitors and zero-knowledge proofs. 
Our approach enables the generation of machine-verifiable, privacy-preserving and tamper-evident SLA claims. 
This removes undesirable dependencies on third-party SLA monitor providers and enhances automated dispute resolution, addressing emerging requirements in cyber insurance.
Our contributions are:
\begin{itemize}
  \item A trusted monitoring architecture and model that collects Service Level Indicator (SLI) data via active and passive probes inside Trusted Execution Environments (TEEs);
  \item A method for converting SLA predicates into verifiable Zero Knowledge Virtual Machine (zkVM) programs with publicly auditable semantics;
  \item A prototype implementation and evaluation showing scalability to over 1M events per hour with near constant-time proof verification of single violation.
\end{itemize}

The remainder of the paper is organized as follows. \Cref{sec:background} reviews the background and related work in SLA monitoring, trust models, and verifiable computations. \Cref{sec:model} defines our monitoring model, its trust assumptions and security properties. \Cref{sec:sys_arch} presents the architecture and verifiability syntax. \Cref{sec:sys_impl} describes our implementation, and \Cref{sec:evaluation} evaluates its performance. \Cref{sec:discussion} reflects on implications and future directions. \Cref{sec:conclusion} concludes.

\section{Background and Related Work}
\label{sec:background}
\newcommand{\tech}[1]{\vspace{0.25em}\noindent{}{\textbf{#1:}}}

This section introduces TEEs for secure SLI measurement and ZKPs for verifiable SLO evaluation, as enablers of integrity, authenticity, and validity. These properties will be formally defined later. It then reviews related work and the state of the art in SLA monitoring.
This aligns with emerging paradigms like DevSecOps, VeriDevOps, or TrustOps~\cite{sadovykh2021veridevops,brito2024trustops}, which advocate continuous, evidence-based trust assessment across cloud infrastructures and stakeholders, now being extended to SLA compliance during service delivery.

\tech{Trusted Execution Environments} TEEs provide hardware-backed mechanisms to isolate computation and protect integrity in untrusted environments~\cite{munoz2023survey,drean2024teaching}. Modern TEEs operate at different levels: Intel TDX and AMD SEV-SNP provide VM-level isolation with encrypted memory, while Intel SGX offers process-level enclaves. TPMs complement these with persistent integrity guarantees via Platform Configuration Registers (PCRs)~\cite{munoz2023survey,drean2024teaching}. Remote attestation enables cryptographic verification of TEE state through reports containing: (i) binary measurements (e.g., MRTD, MRENCLAVE), (i) runtime configuration (e.g., RTMR/PCR values), (iii) ephemeral public keys, and (iv) manufacturer signatures. During execution, TEEs produce supplementary reports with dynamic measurements, enabling continuous integrity verification. They are deployed in the context of cloud platforms~\cite{segarra2024serverless}, trusted compute units~\cite{castillo2025trusted}, confidential containers~\cite{brasser2022trusted}, and trustworthy pre-processing of IoT data~\cite{heiss_2021_ICSOC}. However, integrating them into verifiable SLA monitoring contexts remains an open challenge.

\tech{Zero-Knowledge Proofs} ZKPs enable cryptographic verification of statements without revealing underlying data. Zero-knowledge scalable transparent arguments of knowledge (zkSTARKs), a type of ZKP, provide post-quantum secure, transparent proofs, to ensure computational integrity in outsourced computations~\cite{thaler2022proofs}. A particular application of zkSTARKs are zkVMs, like RISC0 and SP1, which compile programs, for example, to RISC-V bytecode, and generate non-interactive proofs binding: (i) bytecode hash, (ii) public inputs, and (iii) outputs~\cite{lavin2024survey}. They provide self-contained cryptographic artifacts that can be verified by any stakeholder without rerunning the computation or accessing raw data~\cite{ernstberger2024you}. However, proof generation incurs significant overhead (orders of magnitude slower than native execution~\cite{heiss_2021_ICSOC})), prompting optimistic approaches where proofs are generated only for violations, disputes, or audits. This makes cryptographic verification practical in SLA monitoring, where continuous proofs are costly but strong guarantees remain essential. %

\tech{SLA Monitoring} Service Level Agreements (SLAs) define cloud providers’ commitments to performance (e.g., availability, latency) and security (e.g., data protection)~\cite{hussain2014maintaining}, based on measurable SLIs (e.g., request latency) and SLO thresholds (e.g., 95\% of requests under 300ms)~\cite{nicolazzo2024service}. SLA monitoring aggregates low-level metrics into compliance checks~\cite{petcu2014towards}, promoting trust through transparency and predefined escalation paths.
In practice, monitoring is typically provider-controlled~\cite{sghaier2023review}, creating conflicts of interest. Third-party trust delegation~\cite{schubert2018trustworthy} can reduce bias but depends on available and impartial arbiters~\cite{ziegler2023cloud}. Probabilistic detection~\cite{wang2024veriedge} offers another route, though still limited. These shortcomings are critical in domains like cyber insurance, where unverifiable SLA violations impede underwriting, compliance validation, and claims processing~\cite{tsohou2023cyber,aziz2020systematic}.
Trusted compute units~\cite{castillo2025trusted} offer verifiable computation, but target composition correctness rather than SLA evaluation. Economic incentives further complicate trust: modern workflow schedulers often maximize profits by tolerating SLA penalties~\cite{yang2022dual,shen2024cost}. Efforts like rSLA~\cite{ludwig2015rsla}, SLA4OAI~\cite{gamez2019automating}, and blockchain-based models~\cite{weerasinghe2023proof} attempt to address these challenges, but issues remain around data provenance~\cite{kapsoulis2021reinforcing}, measurement integrity~\cite{khan2022blockchain}, and audit-related privacy loss~\cite{alzubaidi2023blockchain}.

\section{Trustworthy SLA Monitoring Model}
\label{sec:model}
\begin{figure}[ht]
    
    \vspace{-1em}
    \setlength{\abovecaptionskip}{2pt}

    \centering
    \includegraphics[width=0.8\textwidth]{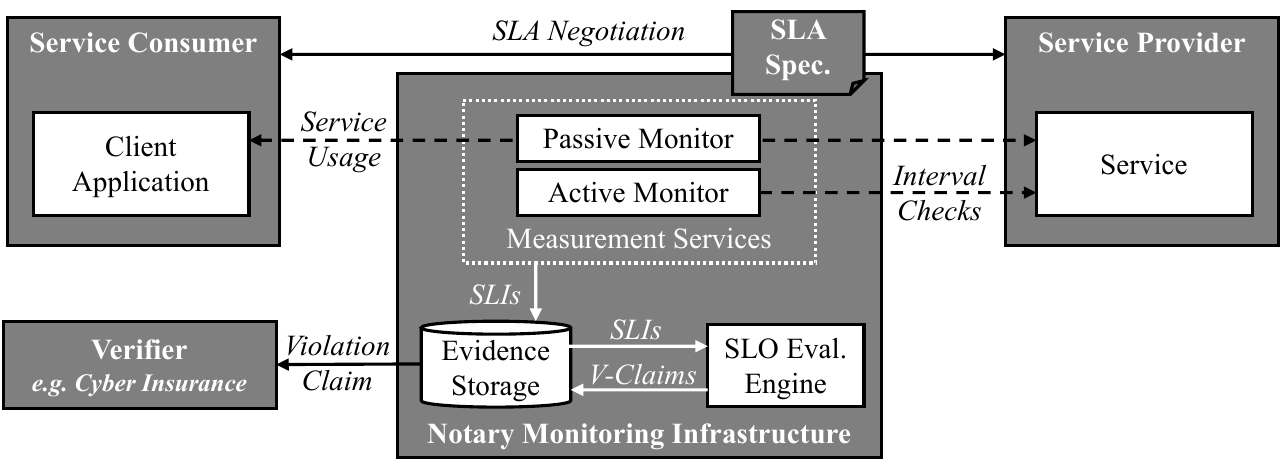}
    \caption{SLA Monitoring Model based on~\cite{keller2003wsla} with Passive and Active Monitor.}
    \label{fig:tw-model}

    \vspace{-1em}

\end{figure}

To set the scene, we introduce a monitoring model that builds upon established approaches to SLA monitoring including the long standing WSLA~\cite{keller2003wsla} and more recent surveys~\cite{nicolazzo2024service}. 
This model, depicted in \Cref{fig:tw-model}, allows us to further refine the problem of trustworthy SLA violation determination. %

\subsection{Roles, Components, and Stages}
The \textit{Service Provider} delivers \textit{cloud services} under the promise of infrastructure reliability and SLA compliance. 
The \textit{Service Consumer} utilizes these cloud resources, applications, or functionality and expects the agreed performance standards. 
Provider and consumer determine their SLA in an \textit{SLA Specification} that defines \textit{Service Level Objectives} (SLOs). 
\textit{Monitors} (or Measurement Services~\cite{keller2003wsla}), either passive or active systems, capture \textit{Service Level Indicator} (SLI) measurements which are persisted in an \textit{Evidence Storage}. 
An \textit{SLO Evaluation Engine} (SEE) aggregates SLIs into \textit{Violation Claims} to check and disclose the violation of SLOs.
To ensure trustworthiness of SLIs and violation claims these components can be deployed on an independent \textit{Monitoring Infrastructured} controlled by a trusted third party (TTP) acting as \textit{Notary}.
Further components described in WSLA~\cite{keller2003wsla}, e.g., SLA Establishment, Deployment, or Management Services, are neglected and not in the scope of this paper. 
These roles and components interact in different stages throughout the SLA lifecycle:
\begin{enumerate}
    \item \textit{Negotiation}: SLA is negotiated and signed by the contractual parties.
    \item \textit{Deployment}: SLA is validated and distributed to the involved components.
    \item \textit{Monitoring}: Monitors collect SLIs, SEEs compute SLA claims. 
    \item \textit{Evaluation}: SLA claims are validated against SLOs to decide SLA violations. 
    \item \textit{Termination}: SLA is terminated based on agreed on conditions. 
\end{enumerate}

These stages directly derive from~\cite{keller2003wsla}, except for \textit{four} which we repurposed to explicitly include SLA evaluation thereby neglecting corrective measures.

\subsection{Problem Refinement}
The trustworthiness of the SLIs and SLA claims are fundamental to guarantee SLA compliant service provisioning and prevent costly disputes between service consumer and provider. 
In the model presented, consumer and provider fully rely on the trustworthiness of the Notary's independent monitoring infrastructure, expecting monitors, evidence storage, and the SEE to be uncompromisable and functioning correctly. 
However, trusted third parties like the Notary are often not available in practice or their contractual terms introduces undesirable dependencies and costs. 
Alternative deployments of such components at the consumer or provider sides, however, lead to information asymmetry that can be exploited by each party.
Given an economic relationship, all parties, including the Notary may manipulate the SLIs and/or SLA claims to maximize their utilities.
Malicious providers may try to conceal SLA violations, consumers might falsely claim violations, and even the Notary may collude with one of the parties, or sloppily neglect security and safety measures of its infrastructure, allowing for external attacks or failures. 
These situations may cause unacceptable financial losses. 
To prevent this, the trustworthiness of the SLIs and SLA claims must be guaranteed in the service ecosystem. 
We consider this to be achieved if all of the following properties of the involved components are met:
\begin{enumerate}
    \item \textit{Integrity}: The monitors and SEE collect SLIs and compute SLA claims correctly ("doing the things right").
    \item \textit{Authenticity}: The collected SLIs and SLA claims are unmodidied and orignate from the expected monitors and SEE. 
    \item \textit{Validity}: The monitors and SEE are running according to their specifications ("doing the right thing"). 
\end{enumerate}
These properties form design objectives which collectively ensure that evidence of SLA compliance or violation is trustworthy and verifiable by all parties.

\section{System Design}
\label{sec:sys_arch}

This section first provides an overview of our verifiable SLA monitoring system design, then presents the key components in more detail. 

\subsection{System Overview}
\begin{figure}[ht]
\setlength{\abovecaptionskip}{2pt}

    \centering
    \includegraphics[width=1\textwidth]{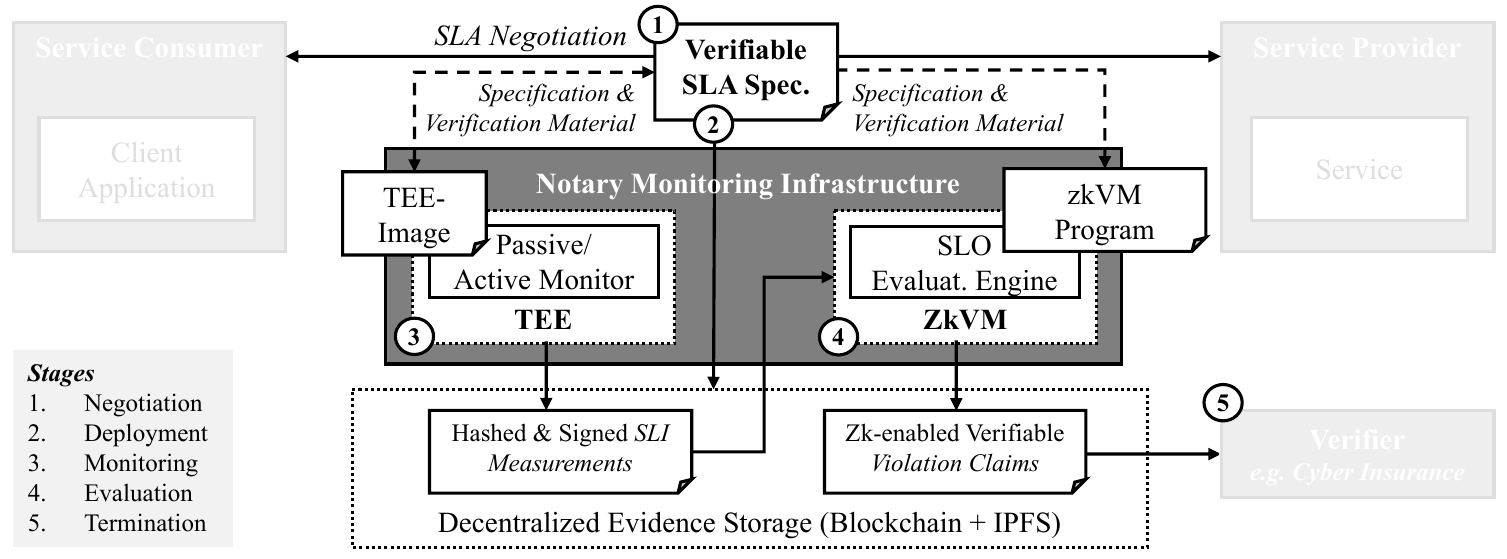}
    \caption{SLI Measurements flow into verifiable SLA compliance claims.}
    \label{fig:sli-flow}

    \vspace{-1.5em}
    
\end{figure}

We address previous design objectives by a) leveraging TEEs for secure monitoring, b) a blockchain-enabled storage system for secure evidence persistence, and c) ZKPs for verifiable SLA claim validation.  
The TEE's hardware-secured isolation prevents manipulation attempts from the monitoring infrastructure provider resulting in trustworthy SLI measurements. 
The ZK-enabled \textit{SLO Evaluation Engine} (SEE) guarantees that the SLIs are correctly checked against the SLOs as determined in the agreed-on SLA. 
A decentralized storage system realizable through blockchains and content-addressable storage systems, like IPFS, ensures the tamper-resistance and accessibility of the SLI measurements and violation claims. 
We describe the interplay and integration of these technologies following the SLA lifecycle stages presented in previous section, as shown in \Cref{fig:sli-flow}. 

\noindent
\tech{1. Negotiation}
Consumer, Provider, and Notary negotiate a \textit{Verifiable SLA Specification} (VSLAS) that extends existing standards through the specification of (1) TEE-enabled \textit{Monitors} and (2) ZK-enabled SEE. 
For (2), the verification key added is cryptographically bound to the implemented SLA terms. 
For (1), the specification is confirmed and verification material is added during deployment. 

\tech{2. Deployment}
The TEE-enabled monitors and the ZK-enabled SEE program are deployed. 
For now, we assume their deployment at the Notary's monitoring infrastructure. 
Once the TEE is booted, an \textit{attestation report} is created and submitted to the decentralized storage system. 
Contained verification material is added to the VSLAS.

\tech{3. Monitoring}
The TEE-based monitor runs at the Notary and collects measurements about the service. 
The measurements are collected in batches.
Merkle Tree-based commitments are created over the batches and cryptographically signed before they are persisted in the evidence storage. The monitor is a passive if it is only monitoring traffic as a proxy between the client and the provider, or it is active if it works as a probe doing independent regular checks to the service.

\tech{4. Evaluation}
Once a violation is detected, the SEE, as the prover, creates a cryptographic proof based on the relevant SLI measurements and the encoded SLA logic.
The proof is submitted to the decentralized evidence storage accessible to contractual parties and available for independent verification of validity, integrity, and authenticity using the corresponding verification key%

\tech{5. Termination}
The ZK-enabled Verifiable Violation Claim can be independently verified using the verification material contained in the VSLAS by the Verifier, e.g. Arbitrator or Cyber Insurance. 

\subsection{Verifiable SLA Specifications and SLO Evaluation} 
The VSLAS extends existing formats through additional specifications and corresponding verification material for both the TEE-based monitors and the ZK-enabled SEE.  
While the former guarantee integrity of the monitoring software's execution and authenticity of the measurements, the latter makes sure the SLA Violation Claims are correctly computed based on the right measurements. 
The VSLAS formalizes SLA predicates as specifications that are translated into provable and publicly auditable ZK-enabled programs, the zkVM bytecode. %
As depicted in Listing~\ref{lst:psla-example}, our specification extends OpenSLO\footnote{https://github.com/OpenSLO/OpenSLO} with named verifiability semantics.
The translation from SLA parameters into a deterministic ZK-proving program can be verified before deployment, allowing all parties to audit the exact logic before agreeing to the SLA. 
This separates what is evaluated (validity) from how it is evaluated (integrity), both verifiably guaranteed.

The \textit{zkVM program} is used by the Verifiable SLO Evaluation Engine (V-SEE) to generate proofs of SLO violations.  
It takes relevant measurements from the TEE-enabled monitors as inputs and executes two key operations: First, it validates the measurements' authenticity using the verification material of the TEE-enabled monitor. 
Second, it aggregates the measurements into SLIs and checks them against predefined SLOs, e.g., through range proofs and predefined thresholds. 
If the measurements' authenticity is proven and the SLIs violate an SLO, an SLA violation is determined.
The resulting ZKP can be verified with corresponding verification key that is cryptographically anchored into the SLA program preventing malicious evaluators to impair the computation's integrity.

By encoding SLA compliance as verifiable predicates, the framework supports higher levels of automation. 
Proof generation can be integrated into orchestration pipelines, producing ZK claims that are independently verifiable, asynchronously by any stakeholder (e.g. automated agents or legal arbitrators), without revealing telemetry or relying on trusted infrastructure. 
This enables scalable, confidential SLA enforcement while minimizing trust and coordination overhead.

\begin{minipage}[t]{0.90\linewidth}
\centering
\begin{lstlisting}[style=psla, caption={OpenSLO extension specification: Adding verifiability.}, label=lst:psla-example]
## STANDARD USE OMMITED BECAUSE OF SPACE
  # VERIFICATION EXTENSION
  verification:
    monitors:
      - type: tee-passive           # TEE-based monitor type
        location: us-east-1         # Deployment location
        tee-provider: intel-tdx     # Hardware attestation
    evidence:
      storage-type: ipfs-private    # Decentralized storage
      blockchain:
        network: base               # Blockchain for anchoring
        contract: "0x..."           # SLA Evidence Registry contract
    aggregation-engine: risc0       # zkVM for proof generation
\end{lstlisting}

\vspace{-1.5em}

\end{minipage}

\subsection{Trusted Monitors and SLIs}
The SLA monitors run inside a TEE to protect their computational integrity. 
Monitors are initialized with specifications for their internal monitoring software co-signed by the provider and consumer. 
Upon boot, the TEE-based monitor registers its hardware-bound public key with the Evidence Storage and exposes a remote attestation endpoint. 
The TEE's key certificates and attestation measurement hash (quote) contained in the attestation report enable the verification of the computational integrity of the TEE, i.e., the right monitoring software is running inside the TEE. 
The TEE's signing key is integrated into the VSLAS of the ZK-enabled SEE to enable proving the authenticity of SLIs collected inside the TEE.  
If the contained TEE reference corresponds to the one in the SLA, validity is assured. 
During monitoring, collected measurements, e.g., latency or response code, are signed and sent to storage, forming the leaves of a Merkle tree. 
The root is signed and checkpointed in the decentralized evidence storage ensuring that all SLIs are accounted for when computing SLOs.

To prevent manipulation of the SLI measurements, storage must be tamper-resistant and equally accessible to all parties. 
To realize such a decentralized storage systems, we leverage blockchain to store non-disclosing commitments and a content-addressable distributed storage system like IPFS to store the corresponding, cryptographically linked measurements~\cite{benisi2020blockchain}. 
Signed measurement data is stored off-chain and linked via cryptographic hashes on-chain, in a SLA Evidence Registry smart contract, enabling scalable retention without incurring blockchain overhead. 
Peer-to-peer replication ensures resilience, while content-addressed identifiers make tampering detectable. 
This combination of decentralized storage and blockchain anchoring is an established pattern~\cite{zhou2019blockchain} that ensures verifiable, persistent, and trustworthy access to SLA evidence.

\section{Technical Realizations}
\label{sec:sys_impl}

We now detail the system implementation\footnote{Repository: https://github.com/ferjcast/Verifiable-SLAs/}, as per the system design in \Cref{sec:sys_arch}.  %

\subsection{Trusted Monitors Implementation: Active and Passive Monitor}

Both monitors rest on a small Monitor Core that executes inside a TEE, like an Intel TDX enclave, deployed in the Notary infrastructure, for instance, with a decentralized TEE provider, to avoid collusion with any party.

At the enclave’s first boot, the Core halts until it receives a co-signed JSON configuration produced during SLA negotiation.
This document names the target endpoints, the measurement interval $\Delta t$, and the schema version. This configuration, co-signed during SLA negotiation, remains immutable throughout the monitor lifecycle. SLA amendments require deploying a new monitor instance with fresh cryptographic keys, ensuring clean security boundaries between different SLA versions. The enclave then produces a remote attestation quote that cryptographically binds the public key to the specific measurements of the initial memory state, the mrTd field of the quote, proving that the key was generated within the authentic monitoring software. This attestation quote, along with the public key and monitor metadata, is submitted to a \textit{SLA Evidence 
Registry smart contract} through a registration transaction signed by both the service provider and consumer.
The Core also exposes an HTTPS \textit{/attestation} endpoint that returns the enclave’s most recent TEE quote; verifiers can thus confirm the initial memory state and runtime measurements directly with remote attestation.

Failure to publish two consecutive batches raises an on-chain heartbeat alert, signaling that the VM was firewalled or powered down; either party can corroborate this by inspecting the project’s audit trail.

\newcommand{\implementation}[1]{\vspace{0.25em}\noindent{}\textbf{#1:}}

\implementation{Active monitor}
Running on the Core, the monitor acts as a probe and, for example, emits a synthetic GET to \textit{/health} endpoint, every $\Delta t$ seconds. It records latency and status, hashes the sample into an in-memory Merkle tree, and, after a configurable records, e.g. 1024, seals the batch to Evidence Storage and signs the Merkle root.

\implementation{Passive monitor}
A lightweight Flask proxy, linked against the same Core, sits inline on the service path and derives latency and status directly from production traffic.
Since it depends also on the amount of requests made to the service, it can be configured to produce batches in time windows when there are measurements.

\subsection{Evidence Storage: Blockchain and IPFS Architecture}%
The Evidence Storage component implements a hybrid architecture combining a permissioned blockchain network with a private IPFS deployment for decentralized content storage. %
The\textit{ SLA Evidence Registry smart contract} implementation defines three primary data structures. The MonitorRegistry mapping associates TEE-generated blockchain addresses with their attestation quotes and operational status. The EvidenceBatch encodes the Merkle root, submitter address, Unix timestamps for the measurement window, and the batch size. The BatchSequence mapping ensures strict ordering by tracking the expected sequence number for each monitor, preventing evidence omission attacks.

Monitors accumulate CIDs from individual IPFS storage operations into a binary Merkle tree, computing parent nodes as keccak256 (left || right) until reaching the root. This 32-byte root represents a cryptographic commitment of measurements while incurring fixed gas costs per batch submission. Verification of individual measurements requires only log(n) hashes to reconstruct the path from leaf to root, enabling efficient proof generation for specific SLI values.
Evidence availability within the IPFS network relies on controlled replication across consortium nodes. The CID-based addressing enables content retrieval from any authorized node in the network, while the blockchain record provides authoritative proof of when evidence was submitted and by which monitor.

As the monitor stores each encrypted measurement in IPFS and collects the resulting CIDs, it maintains an ordered list that preserves the sequence and structure of the measurement batch. Upon reaching the configured batch size of measurements, the monitor creates a JSON manifest file containing the complete ordered array of CIDs along with metadata including the batch timestamp range, monitor identifier, and schema version. This manifest file is itself stored in the IPFS network and registered in the smart contract as part of the batch, receiving its own CID that serves as a permanent reference to the batch composition.

\subsection{Verifiable SLA Specification to zkVM Bytecode}

Our prototype uses RISC0's zkVM, which requires translating SLA specifications into Rust implementations that are then deterministically compiled to RISC-V bytecode. This toolchain provides dual auditability: stakeholders can inspect both the high-level Rust code implementing the SLA logic, and verify its compilation to RISC-V bytecode. The zkSTARK proof system then ensures this exact bytecode was executed during compliance evaluation, creating an auditable chain from specification to execution proof.

While manual specification-to-Rust translation introduces potential for human error, our auditability framework transforms this from a security risk into a process verification step any party can do. Any translation errors become detectable during pre-deployment audit, and the post-deployment cryptographic proofs guarantee execution of the audited logic.

\subsection{SLO Evaluation Engine with zkVM-based Violation Claims}

The SLIs evidence evaluation for SLOs is implemented as a service whose sole trusted component is a zkVM.  Incoming objects are the batch blobs published to the Evidence Storage by the monitors, and identified by their CIDs, and the accompanying SLI measurements, from the smart contract. The SEE executes four steps:

\implementation{1. Pre-flight validation}
Outside the zkVM, the service (i) fetches the ciphertexts from IPFS, (ii) checks the Ed25519 signature on each measurement, and (iii) recomputes the Merkle root over the telemetry. Only the root, the window parameters $\langle t_\text{start},t_\text{end}\rangle$ and the compact metric vector $V$ are passed to the proving environment; in case of giving a wrongful input, the program fails to continue with the next step.

\implementation{2. Proving inside the zkVM}
The compliance predicate, translated from the specification to Rust code, and compiled to use 32-bit RISC-V instruction set, runs in a zkVM STARK proving system, e.g. such as RISC0 or SP1.  For the exemplar SLO “$95\%x$ of latencies below $300\text{ms}$”, the circuit consumes the SLI measurements in the time window under evaluation, and outputs the compliance evaluation, the Monitor's signature verification, and the Merkle tree hash match verification in a single bit $\mathsf{ok}\in\{0,1\}$, for each one of them.  %

\implementation{3. Claim emission and storage}
The zkVM returns $\pi$, $\text{ok}$, and the public inputs $\{\text{root},t_\text{start},t_\text{end}\}.$  These values are wrapped into the standardised claim schema of Listing~\ref{lst:psla-example}, yielding a v-claim. The aggregation execution returns the v-claim and pushes it, together with $\pi$, to Evidence Storage; the corresponding CID and Merkle root are anchored on chain exactly as for monitor batches. Because the proof is non-interactive and succinct, any stakeholder can verify the claim without access to telemetry or to the SEE itself.

\implementation{4. Programmability and upgrade path}
A new or updated SLA predicate is compiled to RISC-V byte-code and hot-swapped by redeploying a container image; no parameter change is required on the monitors. The proving key is deterministically derived from the circuit hash and pinned on chain, ensuring that verifiers always load the correct parameters.%

\section{Evaluation} %
\label{sec:evaluation}

 We evaluate the performance and security properties of our system across three key dimensions: evidence aggregation efficiency, verification cost, and fulfillment of security properties. Experiments simulate TEE-based monitoring and ZKP-based evaluation for a single service.

\subsection{Experiment Design}

Each monitor was deployed on a Phala Network\footnote{https://cloud.phala.network} Confidential VM, as the TEE, instance of type tdx.small with 2 vCPUs and 4 GB of memory, at a cost of 0.138 USD/hr. 
In particular we deploy the passive and active monitor to generate evidence in batches of 512, 1024, 2048, 4096, and 8192 measurements with varying requests per second (rps), with 32, 64, 128, 256 and 512 rps. %
Every batch involved evidence generation, IPFS anchoring, blockchain registration, aggregation, and claim verification. 
We evaluate three proving strategies:  (a) \textbf{Full Batch Disclosure}~\cite{kanwal2015taxonomy} -- All the measurements are disclosed and rely in the monitor's signature; (b) \textbf{Batch-Level Privacy} -- ZKP attests "99.9\% success rate" without revealing measurements, but disputes require disclosing the entire batch; and (c) \textbf{Violation-Level Privacy} -- ZKP proves specific violations (e.g., "measurement \#5,847 exceeded 100ms") with Merkle inclusion, requiring disclosure of only those measurements during disputes while preserving privacy of the remaining measurements in the batch.  %
Blockchain anchoring is performed by submitting evidence CIDs and SHA256 commitments to a smart contract. %

\subsection{Experiment Results}

\begin{figure}[t]
\centering
\setlength{\abovecaptionskip}{2pt}
\begin{minipage}{0.49\textwidth}
\includegraphics[width=0.949\linewidth]{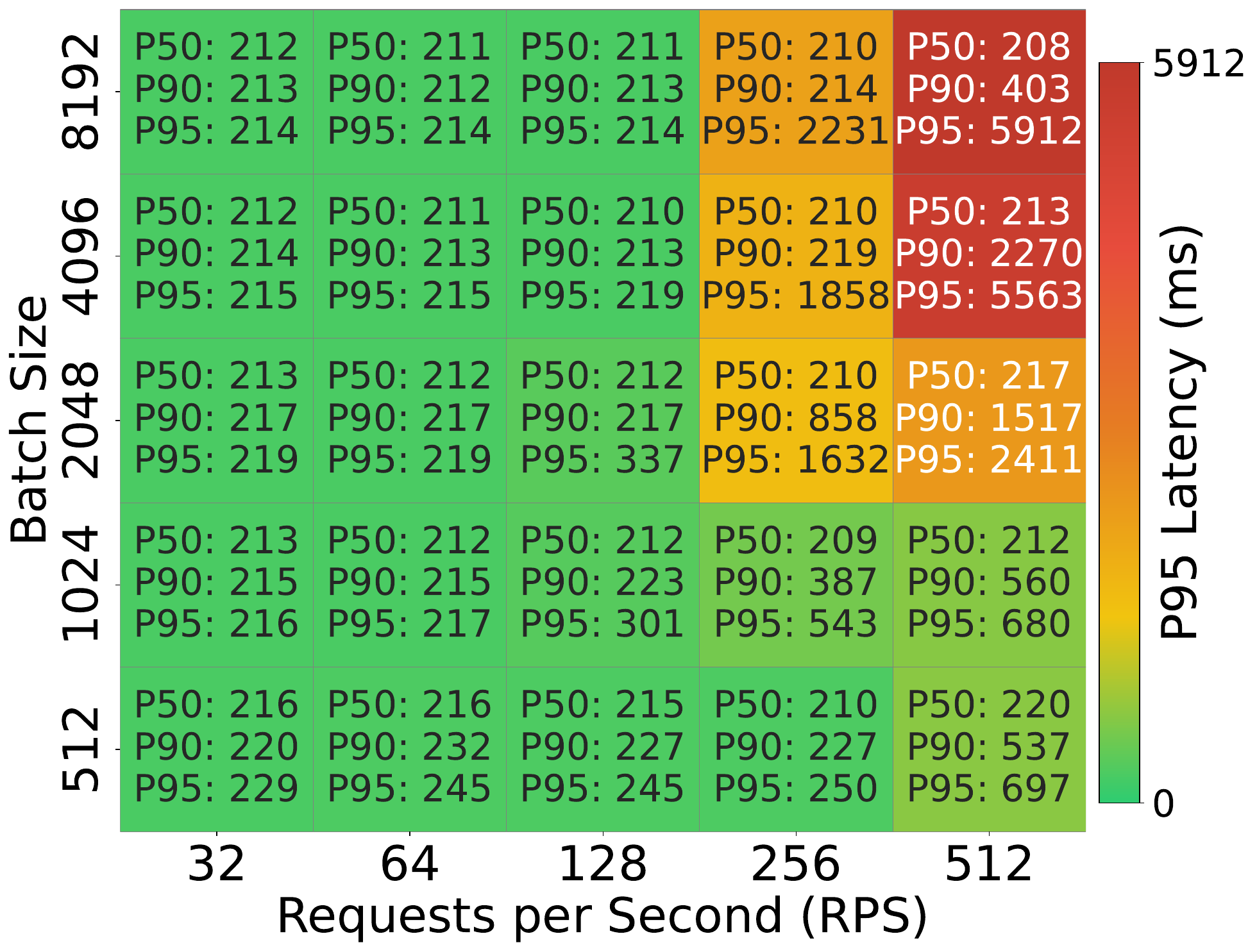}
\caption{Stacked Heatmap for P50, P90 and P95 latency for Batch Size vs RPS.}
\label{fig:heatmap}
\end{minipage}\hfill
\begin{minipage}{0.49\textwidth}
\includegraphics[width=\linewidth]{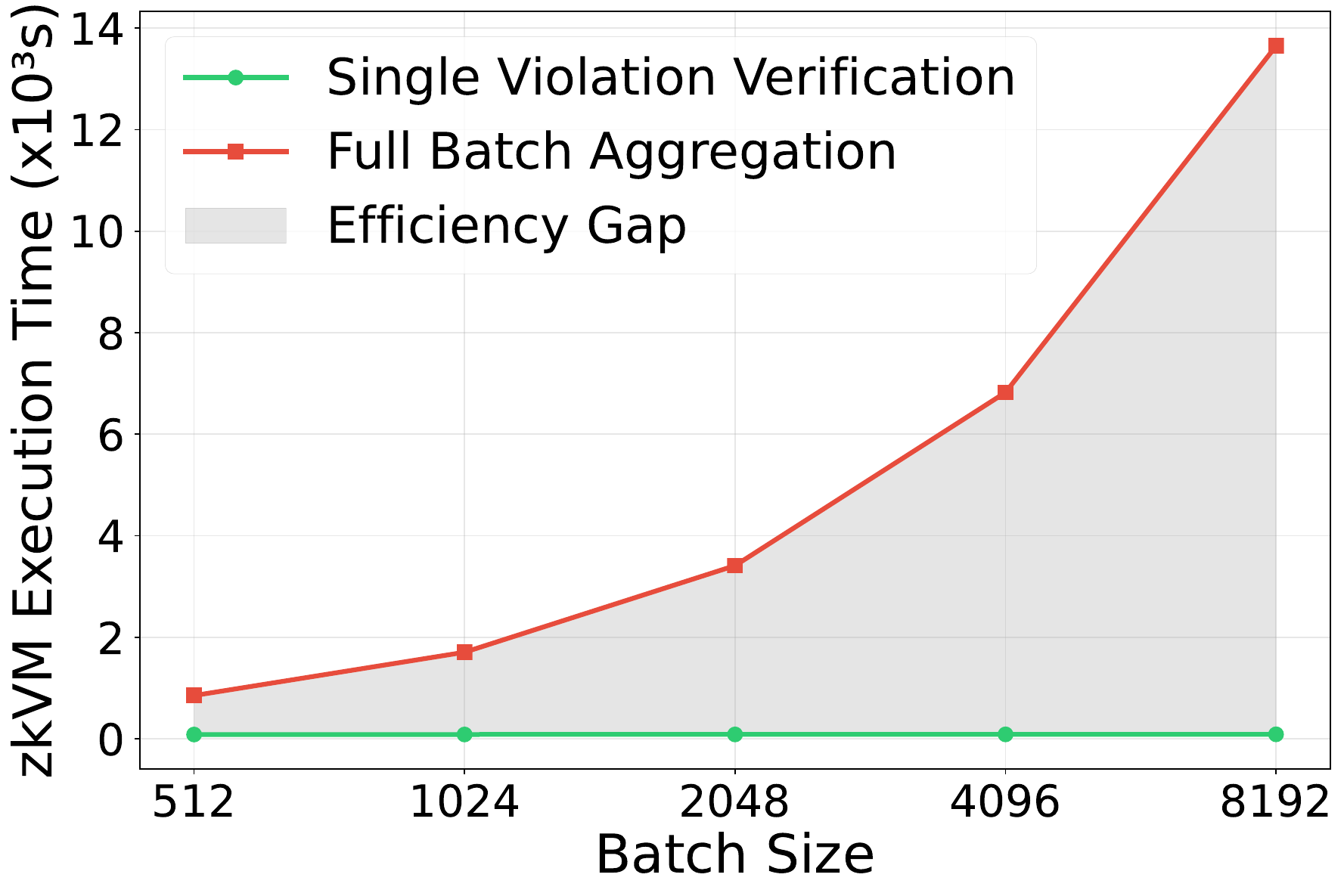}
\caption{Verification time vs number of evidences.}
\label{fig:proving}
\end{minipage}\hfill

\end{figure}

The results are shown in \Cref{fig:heatmap} and \ref{fig:proving}. Considering blocksize limits, gas costs remained constant at 2825k per deployment and 188k for batch anchoring transaction, as only the Merkle root is registered. Since this cost is independent of batch size, the results confirm the scalability of on-chain anchoring with increasing numbers of evidences. %

zkVM aggregation demonstrates consistent performance for single measurement evaluations, requiring only a signature verification and Merkle root recomputation. However, full batch aggregation exhibits linear scaling behavior, where processing time increases proportionally with the number of measurements in each batch, making larger batches computationally more expensive to evaluate. 
Our experiments revealed clear performance boundaries: the monitored service maintained stable operation up to 256 RPS with P95 latencies below the 300ms SLA threshold, but showed initial degradation at 512 RPS. %
Notably, larger batch sizes (2048, 4096, and 8192) significantly impacted latency at both 256 and 512 RPS, demonstrating that batch size becomes a critical performance factor even before reaching maximum system throughput. In contrast, P50 latencies remained relatively stable across all test cases, indicating that the spike degradation occurs specifically when the Merkle tree is sealed, where the process disproportionately affects tail latencies, leaving median performance unaffected.

\section{Discussion and Implications}%
\label{sec:discussion}
Our architecture combines TEEs for secure measurement collection with zkVMs for verifiable compliance evaluation, enabling privacy-preserving SLA claims that are independently verifiable and free from reliance on provider-reported metrics or trusted third parties.
We examined three verification strategies: Full Batch Disclosure, Batch-Level Privacy, and Violation-Level Privacy, each offering a different trade-off between trust properties and performance. Table~\ref{tab:verification-strategies} summarizes their security, verification, and operational properties. While Full Batch Disclosure is lightweight, it depends on trust in the monitor. In contrast, the zkVM-based strategies improve verifiability without revealing full batches.

\begin{table}[htbp]

\vspace{-1em}

\centering
\resizebox{\textwidth}{!}{%
\begin{tabular}{|l||c|c|c|}
\hline
\textbf{Property} & \textbf{Full Batch Disclosure} & \textbf{Batch-Level Privacy} & \textbf{Violation-Level Privacy} \\
\hline\hline
\textbf{Integrity} & TEE signatures &  TEE + zkVM & TEE + zkVM \\
\hline
\textbf{Authenticity} & TEE attestation &  TEE + proof & TEE + proof \\
\hline
\textbf{Validity} & \textbf{Manual inspection} & Bytecode reproducibility &  Bytecode reproducibility \\
\hline
\textbf{Verifier Requires} & Public key + Complete Batch & ZK verifier + public inputs & ZK verifier + public inputs \\
\hline
\textbf{Claim Resolution} & N/A - already public & Reveal entire batch & Reveal only violations \\
\hline
\textbf{Best For} & Public metrics, transparency & Competitive environments & Sensitive operations \\
\hline

\end{tabular}%
}
\caption{Comparison of verification strategies and their trade-offs.}
\label{tab:verification-strategies}

\vspace{-2.5em}

\end{table}

Our architecture balances validity, trust, and cost by combining TEEs and zkVMs. TEEs offer efficient, scalable aggregation with strong integrity and authenticity guarantees, but rely on hardware trust. zkVMs remove the hardware trust assumption and ensure cryptographic validity, though at higher proof generation cost. Nonetheless, this enables a shift from continuous verification to lightweight, on-demand cryptographic guarantees. The design also supports monitor replication and load balancing for linear scalability. %

Direct verification grows linearly with telemetry size (see \Cref{fig:proving}), making it impractical at scale, but aggregation reduces verifier effort to near-constant cost. While zkVM verification time remains stable, proof generation grows with batch size. To reduce overhead, we adopt an \emph{optimistic mode} where proofs are generated only for violations, disputes, audits, or checkpoints, which, in high-uptime scenarios, can drastically cut costs. SLA terms can set thresholds (e.g., only for violations over \$1000), aligning effort with risk. Full Batch Disclosure avoids ZK proof latency but shifts cost to trust. A single monitor key becomes a failure point, requiring audits and legal agreements. These hidden costs often exceed those of cryptographic verification. Using zkSNARKs instead of zkSTARKs could further enable smart contract integration for on-chain SLA enforcement.

Attested TEEs, Merkle commitments, and zkVMs together ensure \textbf{integrity} (correct computation), \textbf{authenticity} (verified origin), and \textbf{validity} (alignment with SLA logic). Merkle trees enable selective audit disclosure without full data exposure. This decentralized auditability reduces operational and legal overhead, resolving the “who monitors the monitors” dilemma through verifiable delegation.

Beyond verification, the system enables SLA-driven automation. Verifiable claims can trigger failovers, renegotiate terms, or automate compensation, supporting SLA-aware orchestration. Even small providers, like micro-clouds or home-labs, can join broader ecosystems by publishing cryptographic proofs.
Standardized, machine-verifiable SLA claims could streamline cyber insurance workflows, like underwriting, claims adjudication, and actuarial modeling. However, current insurer processes lack interoperability with cryptographic claims. Future work should explore formal mappings to legal clauses, machine-readable audits, and trusted risk attestations.
While broader adoption requires research on secure circuit design, predicate verification, TEE support, operational and setup complexity, and legal recognition of cryptographic evidence. These steps would strengthen the foundation for trustworthy, automated service ecosystems.

\section{Conclusion}
\label{sec:conclusion}

We introduced a framework for verifiable SLA compliance claims using TEE-based monitors and zkVM-based aggregation. This architecture ensures integrity, authenticity, and validity of monitoring evidence, enabling privacy-preserving, automated evaluation of SLA conditions.

Our evaluation demonstrates that the system scales efficiently, reduces reliance on trusted third parties, and lowers operational costs by minimizing human involvement in verification and dispute resolution. It also enables selective disclosure for audits and supports automated decision-making in decentralized service environments.
This approach contributes toward more transparent, resilient, and cost-effective SLA management and trustless service ecosystem. It also lays groundwork for emerging applications in cyber insurance, where verifiable compliance evidence could improve underwriting, claims processing, and legal clarity. Future work includes formal verification of predicates, broader TEE support, and exploring the legal standing of cryptographic compliance claims.

\vspace{-1em}

\small{\subsubsection*{Acknowledgements.}
Funded by the European Union (EU) (TEADAL, 101070186). Views and opinions expressed are, however, those of the author(s) only and do not necessarily reflect those of the EU. Neither the EU nor the granting authority can be held responsible for them.
The authors have no competing interests to declare that are relevant to the content of this article.
}

\vspace{-1em}

\bibliographystyle{splncs04}
\bibliography{clean_refs}
\end{document}